\documentclass[sigconf]{acmart}

\AtBeginDocument{%
  \providecommand\BibTeX{{%
    \normalfont B\kern-0.5em{\scshape i\kern-0.25em b}\kern-0.8em\TeX}}}

\copyrightyear{2024} 
\acmYear{2024} 
\setcopyright{rightsretained} 
\acmConference[DIS '24]{Designing Interactive Systems Conference}{July 1--5, 2024}{IT University of Copenhagen, Denmark}
\acmBooktitle{Designing Interactive Systems Conference (DIS '24), July 1--5, 2024, IT University of Copenhagen, Denmark}\acmDOI{10.1145/3643834.3660707}
\acmISBN{979-8-4007-0583-0/24/07}

\usepackage{color}
\definecolor{myred}{rgb}{.8,.0,.0}

\definecolor{myblue}{rgb}{.0,.0,.8}

\newcommand{\recFunctionality}[1]{Enable users to select preferred AI functionality#1}

\newcommand{\recFocus}[1]{Enable users to select radiological findings#1}

\newcommand{\recThreshold}[1]{Enable users to adjust AI thresholds#1}

\newcommand{\recXAI}[1]{Enable users to choose the most suitable XAI method#1}


\begin{document}
\title{"It depends": Configuring AI to Improve Clinical Usefulness Across Contexts}

\author{Hubert D. Zając}
\email{hdz@di.ku.dk}
\orcid{0000-0003-0689-6912}
\affiliation{%
  \institution{University of Copenhagen}
  \streetaddress{Universitetsparken 1}
  \city{Copenhagen}
  \country{Denmark}
  \postcode{2100}
}

\author{Jorge M. N. Ribeiro}
\email{jori@di.ku.dk}
\orcid{0000-0001-5532-2574}
\affiliation{%
  \institution{University of Copenhagen}
  \streetaddress{Universitetsparken 1}
  \city{Copenhagen}
  \country{Denmark}
  \postcode{2100}
}

\author{Silvia Ingala}
\email{silvia.ingala@regionh.dk}
\orcid{0000-0003-2199-385X}
\affiliation{%
  \institution{Rigshospitalet Copenhagen University Hospital}
  \streetaddress{Blegdamsvej 9}
  \city{Copenhagen}
  \country{Denmark}
  \postcode{2100}
}

\author{Simona Gentile}
\email{simona.gentile@regionh.dk}
\orcid{0000-0003-3212-2865}
\affiliation{%
  \institution{Rigshospitalet Copenhagen University Hospital}
  \streetaddress{Blegdamsvej 9}
  \city{Copenhagen}
  \country{Denmark}
  \postcode{2100}
}

\author{Ruth Wanjohi}
\email{ruthwanjohi@nbihosp.org}
\orcid{0009-0005-6210-714X}
\affiliation{%
  \institution{The Nairobi Hospital}
  \streetaddress{Argwings Kodhek Road}
  \city{Nairobi}
  \country{Kenya}
}

\author{Samuel N. Gitau}
\email{samuel.gitau@aku.edu}
\orcid{0000-0002-8908-4735}
\affiliation{%
  \institution{Aga Khan University Hospital}
  \streetaddress{3rd Parklands Ave}
  \city{Nairobi}
  \country{Kenya}
}

\author{Jonathan F. Carlsen}
\email{jonathan.frederik.carlsen@regionh.dk}
\orcid{0000-0002-6724-7281}
\affiliation{%
  \institution{Rigshospitalet Copenhagen University Hospital}
  \streetaddress{Blegdamsvej 9}
  \city{Copenhagen}
  \country{Denmark}
  \postcode{2100}
}

\author{Michael B. Nielsen}
\email{michael.bachmann.nielsen@regionh.dk}
\orcid{0000-0002-9380-1688}
\affiliation{%
  \institution{Rigshospitalet Copenhagen University Hospital}
  \streetaddress{Blegdamsvej 9}
  \city{Copenhagen}
  \country{Denmark}
  \postcode{2100}
}

\author{Tariq O. Andersen}
\email{tariq@di.ku.dk}
\orcid{0000-0002-9342-5520}
\affiliation{%
  \institution{University of Copenhagen}
  \streetaddress{Universitetsparken 1}
  \city{Copenhagen}
  \country{Denmark}
  \postcode{2100}
}

\renewcommand{\shortauthors}{Zając et al.}

\begin{abstract}
Artificial Intelligence (AI) repeatedly match or outperform radiologists in lab experiments. However, real-world implementations of radiological AI-based systems are found to provide little to no clinical value. This paper explores how to design AI for clinical usefulness in different contexts. We conducted 19 design sessions and design interventions with 13 radiologists from 7 clinical sites in Denmark and Kenya, based on three iterations of a functional AI-based prototype. Ten sociotechnical dependencies were identified as crucial for the design of AI in radiology. We conceptualised four technical dimensions that must be configured to the intended clinical context of use: AI functionality, AI medical focus, AI decision threshold, and AI Explainability. We present four design recommendations on how to address dependencies pertaining to the medical knowledge, clinic type, user expertise level, patient context, and user situation that condition the configuration of these technical dimensions.
\end{abstract}




\begin{teaserfigure}
  \includegraphics[width=\textwidth]{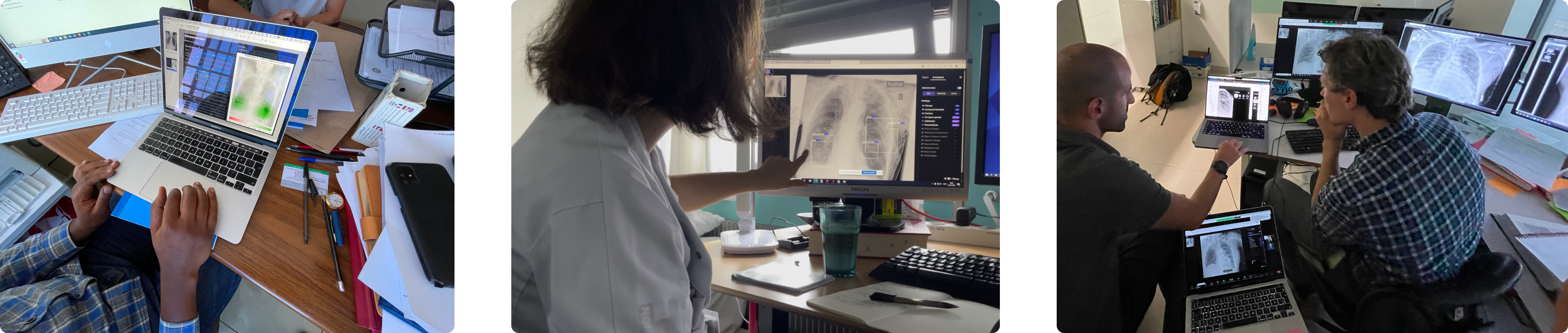}
  \caption{Design interventions in different radiology settings in Kenya and Denmark. Three versions of an AI-based prototype were used to explore configuration opportunities to achieve clinical usefulness across clinical sites (from left: version I, II, II).}
  \Description{A collage of three images showing design sessions and interventions with user interface mock-ups and working prototypes. The images depict various scenarios, including pointing at a laptop screen displaying chest X-rays and explainable AI methods, and collaborative tasks in front of computers.}
  \label{fig:teaser}
\end{teaserfigure}

\maketitle

\section{Introduction}
Artificial Intelligence (AI) models repeatedly match or outright outperform radiologists in narrowly defined detection tasks \cite{Nam2019DevelopmentRadiographs, Ardila2019End-to-endTomography, Rodriguez-Ruiz2019Stand-AloneRadiologists, Rajpurkar2018DeepRadiologists}. There are multiple studies claiming that AI-based systems enhance radiologists' work, either by increasing accuracy or reducing time spent on each examination \cite{Li2021TheReview}. These claims, however, are based on retrospective evaluations conducted in laboratory settings. When looking closer into the state of the art of clinician-facing AI, the claims of utility weaken \cite{Zajac2023Clinician-FacingHCI}. For example, Roberts et al. \cite{Roberts2021CommonScans} found that out of the 62 AI models detecting and predicting COVID-19 on chest X-rays and CT scans that were described in the literature, none were deemed to be useful for clinical purposes. Furthermore, evaluations of the handful of systems approved by the authorities in the United States and European Union \cite{Adams2021ArtificialImages} revealed that their clinical impact when integrated into practice remains mostly unclear \cite{vanLeeuwen2022HowOutcomes, Weikert2020ARadiology.}. A similar study by Lehman et al. \cite{Lehman2015DiagnosticDetection} showed no improvement in patient outcomes after the successful integration of an AI-based support tool for mammography screenings. Strohm et al. claimed that one of the primary causes of AI's lack of success in radiology until now is due to  "uncertain added value for clinical practice of AI applications" \cite{Strohm2020ImplementationFactors.}. What these studies show is that the clinical usefulness of hitherto AI-based support systems is limited.

Researchers with diverse backgrounds (AI, Health, and Human-Computer Interaction (HCI)) investigated what makes AI-based support systems clinically useful. Based on the previous work, we define clinical usefulness as the overarching quality of AI-based support systems emerging from the interplay of their real-world performance, clinical efficacy, local applicability, and end-user acceptance in a situated clinical context for concrete end-users. First, robust performance in real-world settings is essential, as subpar performance has been found to increase workload and disrupt clinical routines \cite{Zajac2023Clinician-FacingHCI, Wang2021BrilliantDeployment, vanLeeuwen2021ArtificialEvidence.}. Second, the evaluations, primarily assessing technical performance metrics through randomised clinical trials (RCTs), must encompass tangible clinical outcomes and patient benefits. Health researchers have been advocating for more flexible assessment methodologies aligned with the iterative nature of AI deployment \cite{Liu2019ReportingNeeded, Berkovsky2023MovingAI, Kelly2019}. Third, end-user acceptance, supported by qualities like trust and usability, emerges as pivotal for successful use in clinical practice \cite{Burgess2023HealthcareTrust, Cai2019Human-CenteredDecision-Making, Jacobs2021DesigningLens}. Altogether, for an AI-based system to be clinically useful, it must perform well, benefit patients, and be accepted by clinical end-users working in different clinical contexts.

In this paper, we investigate \textbf{how to design AI for clinical usefulness in different clinical contexts.} This study was conducted as a part of a larger research and development project focused on innovating an AI-based system to assist examinations of chest X-rays in Denmark and Kenya. Here, we define innovation as the entirety of work conducted to create an AI-based system, from creating the datasets the AI is trained on through design and development to its integration and use in practice. We conducted 19 design sessions and design interventions (online and collocated) with 13 radiologists from 7 clinical sites in Denmark and Kenya. Throughout the design study, we explored a range of user interface mock-ups and three versions of a web-based prototype of an AI-based support system with prioritisation and decision-support functionalities.

We conceptualised four technical dimensions of radiological AI support that need to be configured to maximise its clinical usefulness. The technical dimensions uncovered through the design interventions span \emph{AI functionality, AI medical focus, AI decision threshold, and AI Explainability}. These decisions constitute the critical aspects of radiological AI-based support systems and must be configured in relation to the local social dimensions of clinical AI. 

Moreover, to support configuration during innovation, we deconstructed social dimensions, conditioning how each of the technical dimensions supports final clinical usefulness. Namely, how \emph{medical knowledge, clinic type, user expertise level, patient context}, and \emph{user situation} affect the clinical usefulness of the technical dimensions. 

Finally, we discuss how these dependencies should be accounted for throughout the innovation processes to successfully configure future systems before-use and enable meaningful configuration in-use. Based on the design interventions, we offer four concrete design recommendations addressing the configuration needs of each of the conceptualised technical dimensions of clinical AI.
\section{Related Work}
\subsection{Clinical Usefulness of AI Systems in Healthcare}
The hitherto evidence of AI's positive influence on clinical practice is limited \cite{Tariq2020CurrentEvidence, vanLeeuwen2021ArtificialEvidence., Mehrizi2023How2021.}. Research on the real-world effect of AI in healthcare tends to be discrete and focusing on confined goals \cite{Zajac2023Clinician-FacingHCI}. However, to provide clinical value AI-based systems have to dovetail contributions from Human-Computer Interaction, AI, and Health into a cohesive vision \cite{Zajac2023Clinician-FacingHCI, Wang2021BrilliantDeployment, Fogliato2022WhoImaging}. 

First, clinical usefulness necessitates robust performance \cite{Zajac2023Clinician-FacingHCI}. This primarily has to be true in real-world settings, retrospective evaluations in lab environments do not speak to the final performance of a system. For example, in a real-world evaluation of an acclaimed ML model for detecting diabetic retinopathy, 21\% of all cases were deemed ungradable \cite{Beede2020}. Poor performance also leads to increased workload \cite{Romero-Brufau2020ASupport, Wang2021BrilliantDeployment, Petitgand2020InvestigatingStudy}, additional time spent on discerning false positive predictions \cite{Sendak2020, Sendak2020a}, or breakages to work routines \cite{Gastounioti2014CAROTIDPatients}. Van Leeuwen et al. \cite{vanLeeuwen2021ArtificialEvidence.} reported that out of 100 CE-approved radiological AI-based systems, 64 showed no peer-reviewed evidence of clinical efficacy. Most evidence for the remaining 36 systems focused on diagnostic accuracy, not real-world clinical outcomes.

Second, clinical usefulness necessitates clinical efficacy \cite{Kelly2019}. However, randomised clinical trial (RCT) - a focused, systematic, rigorous, and insulated method commonly used to evaluate the validity of clinical interventions independent of external confounders - is often following the traditional sequential paradigm of work characteristic for drug development \cite{Blandford2018SevenInterventions}. In this tradition, the intervention is evaluated only when deemed complete \cite{Campbell2000FrameworkHealth}. When translating this mentality to AI-based systems, not only does it hinder innovation, but it also results in the evaluation of AI through the measure of technical performance \cite{Liu2019ReportingNeeded}. While technical performance is the backbone of useful AI, clinical efficacy is not its immediate consequence \cite{Berkovsky2023MovingAI, Shah2019MakingUseful}. For example, Lehman et al. \cite{Lehman2015DiagnosticDetection} conducted a prospective evaluation of a computer-aided detection system supporting mammography reporting. Researchers concluded that the use of AI had no "established benefit to women." Instead, healthcare researchers are opening up towards more flexible evaluation approaches that align with the iterative and situated nature of AI innovation and "go beyond measures of technical accuracy to include quality of care and patient outcomes" \cite{Kelly2019, Cresswell2017DrawingTechnology}. Achieving high performance but in metrics that are clinically relevant is the next step towards clinically useful AI-based systems.

Third, clinical usefulness necessitates clinical organisational acceptance. HCI community's claim to fame is understanding that regardless of a system's performance, it will not have any impact if no one wants to use it. Thus, many facets of making clinical AI an appealing solution were explored. Trust has been hallmarked as a critical quality of clinical AI. HCI researchers investigated its origin \cite{Burgess2023HealthcareTrust, Bach2023IfSupport} and dependencies \cite{Procter2023HoldingHealthcare}, as well as issued recommendations for design \cite{Jacobs2021DesigningLens}. Explainable AI (XAI) has been the most promising answer to enhance trust, support oversight, and increase the perceived usefulness of clinical AI \cite{Cai2019Human-CenteredDecision-Making, Xie2020CheXplain:Analysis, Lee2023UnderstandingMaking, Gu2021LessonsPathologists}. AI as a new source of information and agency prompted the exploration of new ways of reasoning and human-AI collaboration \cite{Berge2023DesigningCalls, Bossen2023BatmanSpecialists, Donahue2022Human-AlgorithmUnfairness, Cai2019a}. Researchers also investigated AI's position in a clinical decision-making process \cite{Kapania2022BecauseIndia} and the rationale behind integration opportunities into clinical practice \cite{Jacobs2021DesigningLens, Yang2016InvestigatingHelp, Sendak2020a, Sandhu2020IntegratingStudy}. They argued that the workflows, current work practices, and the broader sociotechnical context should also be taken into account when implementing clinical AI-based systems \cite{OsmanAndersen2021RealizingWild, Coiera2019TheReality, Jacobs2021DesigningLens, Thieme2020MachineHealth, Robertson2021ModelingAlgorithms, Zajac2023Clinician-FacingHCI}. Addressing these concerns is crucial for AI to have a chance at benefiting patients and being accepted by healthcare professionals. 

Altogether, for an AI-based system to be clinically useful it must perform well, benefit patients, and be accepted by clinical end-users. However, oftentimes the innovation of clinical AI is conducted in silos and the work is not guided by the ultimate goal of clinical usefulness \cite{Zajac2023Clinician-FacingHCI, Blandford2018SevenInterventions}. We need to investigate how AI-based systems can be configured to support these three goals and ultimately result in clinically useful AI.

\subsection{System Configurability}
Configurability has been long considered crucial to the appropriation of IT systems \cite{Kyng1997ComputersContext, dourish1997accounting, Dourish2003TheDocuments}. There are two types of configurability that should be explored in the context of this study: before-use and in-use \cite{hertzum2019configuring}. 

Before-use configurability typically involves the active participation of end-users in the design processes, aiming to tailor systems to their specific needs and preferences \cite{hertzum2019configuring}. Various methods and approaches have emerged to facilitate meaningful engagement with end-users, such as participatory design techniques \cite{Kyng1997ComputersContext}. Acquiring an understanding of work practices and work environment, but also technology aspects of a future system and changes it may introduce, is critical for developing systems that effectively respond to user needs \cite{Kensing1993PD:Toolbox}. This understanding enables developers and designers to implement systems that are not only technically sound but also contextually appropriate.

However, according to Stewart and Williams \cite{Stewart2005TheWT}, the paradigm of user-centred design does not properly answer the challenges of implementing useful systems. Rather, the final usefulness of a system is created iteratively through the acts of in-use configuration. This stance echoes Suchman who recognised the need for design activities to continue after a system's deployment \cite{Suchman.2002.Working}.

The in-use configuration may cover functionalities, user interface, or other settings that let the end-users adjust the system to their preference and work environment  \cite{Wulf2001DirectActivities}. However, the system is not the only configurable arena. The environment also undergoes a process of configuration to the new system. The in-use configuration processes encompass changes to the "technical environment, organisational relations, space technology relations, as well as people’s connections to other people, to other places, and work materials"  \cite{Balka2006MakingWork}. Dourish highlights how the appropriation of IT systems in practice is an act of both adapting the technology and adapting the practices to fit into the new reality \cite{Dourish2003TheDocuments}. 

As usual with AI, the matter of configuration is burdened by the immutability of certain aspects of the system in-use and the dependency of early design decisions on the use context \cite{Zajac2023Clinician-FacingHCI}. HCI researchers investigating the design of AI-based systems learned that it is impossible to envision all aspects of clinical AI-based systems before deployment. As a result, the final capabilities of such systems only take shape after they have been deployed. \cite{Girardin2017WhenScientists, Yang2018InvestigatingLearning}. On the opposite end of AI innovation, i.e., prior to data labelling, Zając \& Avlona \cite{Zajac2023Clinician-FacingHCI} established that very concrete choices and assumptions about the final context of AI use form the data used for AI training and, by extension, shape the space of capabilities of future AI-based systems. This vicious cycle of dependencies prompted researchers into new ways of thinking about AI innovation. Edwards et al. \cite{Edwards2007UnderstandingID} proposed the concept of "growing" to foreground the need for almost organic adoption and adaptation of new IT systems in an existing environment. Elish and Watkins presented a similar argument \cite{Elish.2020.Repairing} who emphasise that early realisation of clinical AI and acknowledgement and support of the necessary "repair work" are crucial to counter the risk of a system remaining "a potential solution", i.e., a solution that is not viable when actually implemented. 

We see the problem of configuration of clinical AI, as a problem of obtaining reliable information related to design decisions made during the innovation process. The emergence and propagation of dependencies (or "sociotechnical interdependencies", see \cite{Zajac2023Clinician-FacingHCI}) at the point of deployment hamper the ability to configure clinical AI-based systems in-use. At this point, the assumptions about the context of use are already ingrained in the AI model. We want to support the configuration of radiological AI-based systems for clinical usefulness by uncovering the dependencies anchored in clinical contexts and linking them with specific design decisions. This extended understanding of contextual factors will allow developers and designers to implement radiological AI support configurable and useful across clinical contexts.

\section{Methodology}
In this paper, we explored how to design radiological AI-based systems for clinical usefulness across contexts. This study was part of a larger project set to design and develop an AI-based support tool for radiologists examining chest X-rays, funded by the Innovation Fund Denmark (0176-00013B). The project is a multidisciplinary collaboration between the Department of Computer Science at the University of Copenhagen, the Department of Radiology at Rigshospitalet Copenhagen University Hospital, and Unumed ApS. Due to the project's goals, the future system should support radiologists in Denmark and Kenya. To take into account the diversity of practices and contexts, we conducted design research in seven different healthcare settings across the two countries (Table \ref{tab:sites}), which included: (1) imaging clinics - where medical imaging services such as chest radiographs, ultrasounds, or CT scans (only K3) are provided to patients referred by external physicians, (2) general hospitals - that offer primary and secondary care and refer patients requiring more specialised care to other facilities, and (3) specialised hospitals - that provide tertiary and quaternary care, handling the most complex medical procedures in their respective countries.

{\small
\begin{table}
    \centering
    \begin{tabular}{cccc}
    \toprule
         \textbf{\#}&  \textbf{Type} & \textbf{Radiologists (Total)} & \textbf{Country} \\
         \midrule
         K1&  Small General Hospital& 1 (<5) & Kenya \\
         K2&  Small General Hospital&  1 (1) &Kenya \\
         K3&  Imaging / Teleradiology Clinic&  1 (1)& Kenya \\
         K4&  Specialised Hospital&  1 (<20)& Kenya \\
         K5&  Big General Hospital&  3 (10) & Kenya \\
         D1&  Specialised Hospital&  6 (>100) & Denmark \\
 D2& Imaging Clinic & 1 (<5) & Denmark \\
 \bottomrule
    \end{tabular}
    \caption{Clinical sites included in the study and the number of radiologists (N=13) participating from each of the sites in relation to the total number of employed radiologists. One radiologist with double affiliation in D1 and D2.}
    \label{tab:sites}
\end{table}
}

The participants were recruited through email and the professional networks of the project members. Nine senior (consultant) radiologists and four junior (in-training) radiologists joined the study (Table \ref{tab:interventions}). Junior radiologists' reports must be most often approved by a senior colleague before sharing with clinicians. The senior radiologist's assessment is final. Participants were not compensated, and we collected written consent from all participants. According to the authors’ institutions’ institutional review boards (IRBs), our study was considered non-interventional and thus exempt from a formal ethical review.

{ \small
\begin{table}
    \centering
    \begin{tabular}{clcccc}
    \toprule
\textbf{\#} & \textbf{Participant}&\textbf{Expertise}&  \textbf{Clinical site} &  \textbf{Length}&  \textbf{Prototype} \\
\midrule
 I01& P01 &Senior&  K1&  60m&  I \\
 I02& P02 &Senior&  K2&  60min&   I\\
 I03& P03 &Senior& K3& 120min& I\\
 I04& P04 &Senior& K4& 50min& I\\
 I05& P05 &Senior& K5& 80min& I\\
 I06& P06 &Senior& K5& 80min& I\\
 I07& P07 &Senior& K5& 60min& I\\
 S08& P08 &Junior& D1& 70min& II\\
 I09& P09 &Junior& D1& 50min& II\\
 I10& P10 &Senior& D1& 30min& II\\
 I11& P11 &Senior& D1 / D2& 30min& II\\
 S12& P08 &Junior& D1& 60min& II\\
 S13& P10 &Senior& D1& 30min& III\\
 S14& P12 &Junior& D1& 80min& III\\
 I15& P11 &Senior& D1 / D2& 45min& III\\
 I16& P10 &Senior& D1& 30min& III\\
 I17& P13 &Junior& D1& 40min& III\\
 S18& P05 &Senior& K5& 95min& III\\
 S19& P04 &Senior& K4& 70min& III\\
 \bottomrule
    \end{tabular}
    \caption{Radiology participants that took part in the study. Interventions - I, Sessions - S.}
    \label{tab:interventions}
\end{table}
}

\subsection{Research Through Design: Design Interventions with Working Prototypes}
\begin{figure*}
    \centering
    \includegraphics[width=1\linewidth]{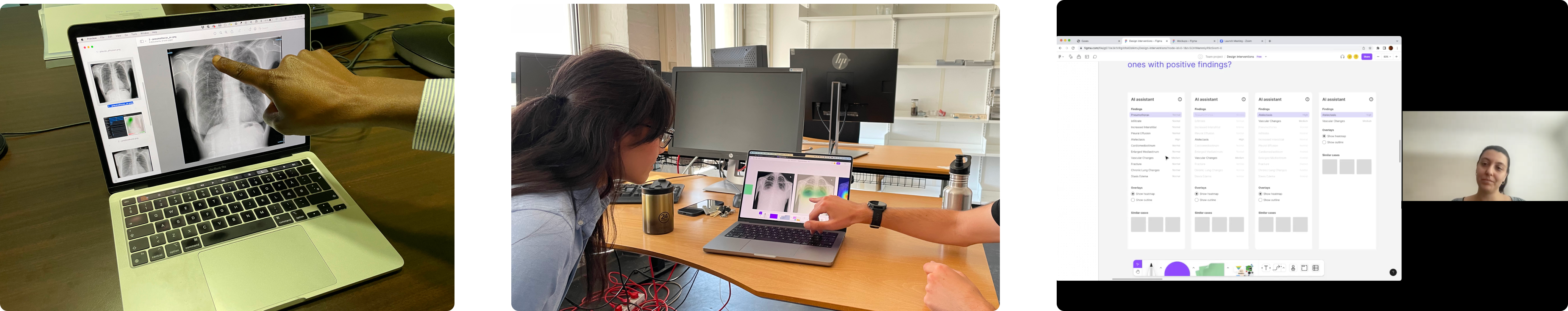}
    \caption{Online and co-located design sessions with user interface mock-ups and functional prototypes (from left: iteration I, II, and III)}
    \Description{A collage of three images showing online and co-located design sessions and interventions with user interface mock-ups and working prototypes. The images depict various scenarios, including pointing at a laptop screen displaying chest X-rays, colourful graphs and data visualisations, and interface mock-ups.}
    \label{fig:interventions}
\end{figure*}

\begin{figure*}
    \centering
    \includegraphics[width=1\linewidth]{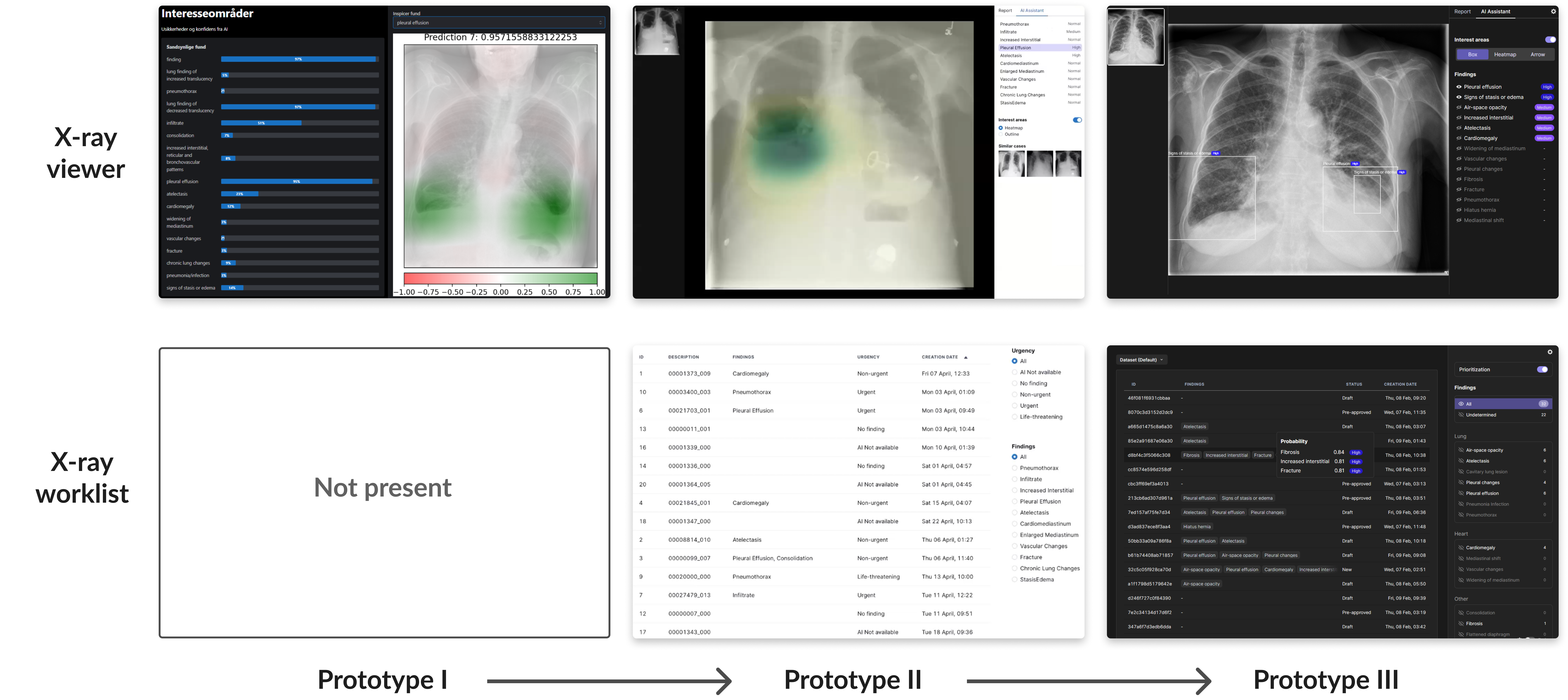}
    \caption{The collage showcases three iterations of the AI prototype, including two screens: the X-ray viewer and the X-ray worklist.}
    \Description{The image depicts six images of three prototypes of an AI-based system for chest X-rays (2 images per prototype). For each prototype, two screens are presented: an X-ray viewer and an X-ray work list view. }
\label{fig:prototypes}
\end{figure*}

To explore the clinical usefulness of AI in different radiology contexts, we undertook a research through design approach  \cite{zimmerman2007research-through-design}. We conducted three iterations based on a series of design sessions and design interventions using mock-ups of user interfaces (Prototype I) and working prototypes (II and III) (Fig. \ref{fig:teaser}, Fig. \ref{fig:interventions}, and Fig. \ref{fig:prototypes}). The three iterations were determined by decisions to deploy major changes in the web-based prototypes, i.e. version 1-3, followed by gathering feedback from the participants. The design sessions were carried out both online and collocated with radiologists in hospital offices. During these sessions, we obtained medical domain knowledge, typically by clarifying questions about radiology work and X-rays, but we also collectively explored the design space through a range of mock-ups and prototypes. The design interventions were carried in-situ with the performative purpose of exploring how the proposed solutions would be enacted close to real-world radiology practices. A design intervention, as defined by Halse and Boffi \cite{Halse2016DesignInquiry}, is a method that integrates design and ethnography and "enables new forms of experience, dialogue, and awareness about the problem to emerge" (see also  \cite{blomberg2013reflections, blomberg1996work-oriented-designproject}). It is an experimental form of inquiry that enables a positioning "in-between what is already there and what is emerging as a possible future" \cite{andersen2011design}. 

In our case, this meant that we intervened in the radiologists' everyday work settings with design artefacts as a vehicle for exploring the dependencies of AI usefulness in situated contexts. During observations of the radiologists' work practices, we brought in the prototypes and mock-ups as a way to enact while experimenting with new forms of AI support in radiology. The benefit of this approach was the possibility to engage radiologists in moving between considering the proposed solutions and envisioning alternatives while constrained by the requirements of the local context. This mode of research was important for this study because it provided more grounded and realistic visions of how AI could become clinically useful across hospital contexts. 

In total, we conducted thirteen design interventions and six design sessions (Table \ref{tab:interventions}) with thirteen radiologists in Denmark and Kenya, lasting between 30 and 120 min (avg. 60 minutes). In between sessions and interventions, we designed a range of user interface mock-ups using Figma, consisting of different AI functionalities and alternatives to interactive features. A total of three versions of a web-based prototype, which included an AI model developed in the greater part of the project. This meant that the participants in this study interacted with real data and real output from the AI model during design interventions. Importantly, the data was completely anonymised, and no other medical information about the patients was available. The mock-ups, prototypes and feedback from the participants became input for multiple design meetings within a group of three of the authors (HDZ, JMNR, TOA) and included high-level discussions with machine learning engineers at Unumed ApS. Here, insights were discussed, and decisions were made regarding what the following design explorations should consist of. All design sessions and interventions were audio-recorded and machine-transcribed to support thematic analysis.

\subsubsection{Prototypes}
As part of the greater project, a deep learning-based model was developed by machine learning engineers at Unumed ApS to detect selected radiological findings \cite{Zajac2023GroundAI}. The AI model was developed using a convolutional neural network. The first prototype was merely a proof of concept, not designed to collect feedback from external domain experts. It was developed to guide future work in terms of model development and data labelling. However, inspired by earlier research \cite{Zajac2023Clinician-FacingHCI, Zajac2023GroundAI}, we considered it an opportunity to engage in more concrete discussions on the merit of clinical usefulness with medical professionals at an early stage of the innovation work.

The second and third iteration of the prototype consisted of an interactive web application designed to emulate a DICOM viewer. The web application integrated with the AI model developed within the bigger project.  This connection enabled us to work with real data and, thus, explore with fidelity the interactions of the radiologists with the system. For the design interventions, radiologists were given access to the prototype, either in-person or remotely. They were requested to choose the next examination to report, following their usual practice and using information displayed in the prototype. Then, they were asked to interpret the selected examination without the use of AI and with AI decision support. Moreover, they were asked to configure the AI tool using available options to fit their practice. Finally, they were encouraged to explore the prototype independently and interact with any element of the user interface.
 
\subsection{Analysis Positionality}
The data analysis was conducted by the first and last authors (HDZ and TOA) with backgrounds in Health Informatics, HCI, and AI (5+ \& 15+ years of experience). Moreover, before the analysis of the data from the design sessions and design interventions, the two co-authors concluded extensive ethnographic investigations into the work practices of radiologists from the visited sites with a particular outlook on opportunities for AI support (described in a manuscript prepared for publication). First-hand experience with the work practices and similarities and differences across clinical settings informed the initial analysis of this data.

\subsection{Data Analysis}
We used reflective thematic analysis \cite{Braun2006UsingPsychology} to analyse collected data (transcriptions of the design interventions). The analysis took place in Dovetail - a web application for qualitative data analysis. Except for the transcription software, no AI-based analysis support was used in this study. The two authors familiarised themselves with the collected data after every iteration of the design sessions and design interventions when deciding on the next focus. Moreover, the two authors, prior to coding, based on their fieldwork experience (60+ hours) and a literature review \cite{Zajac2023Clinician-FacingHCI}, devised three bucket themes to support the later organisation of codes: type of clinical site, domain expertise of medical professionals, and patient and situational context. Additionally, a fourth residual category was added not to limit coding. Next, to test the bucket themes, the two authors coded one transcript each for any references to challenges, preferences, dependencies, and configurations in relation to AI and their clinical practice. After this test, the fourth bucket theme was renamed to technical dependencies. The first author coded the remaining transcripts following the same directions. The two authors met weekly to discuss the coverage of the coding and future conceptualisation of themes. The themes were created within their respective bucket themes based on their grounding in the clinical context. Importantly, the division of codes between the bucket themes was never final and was used only to support analysis of the significant amount of codes (n=260). Through discussion, reflection on data across the interventions, and fieldwork experience, the authors iteratively clarified themes and reorganised data, moving away from the original bucket themes (while maintaining their initial assignment known). This interpretative work was conducted twice, creating ten reflective themes. The ten themes were framed as dependencies conditioning four specific design decisions that formed an AI-based support design space.
\section{Configuring Four Technical Dimensions of Clinically Useful Radiological AI}
\begin{figure*}
    \centering
    \includegraphics[width=0.75\linewidth]{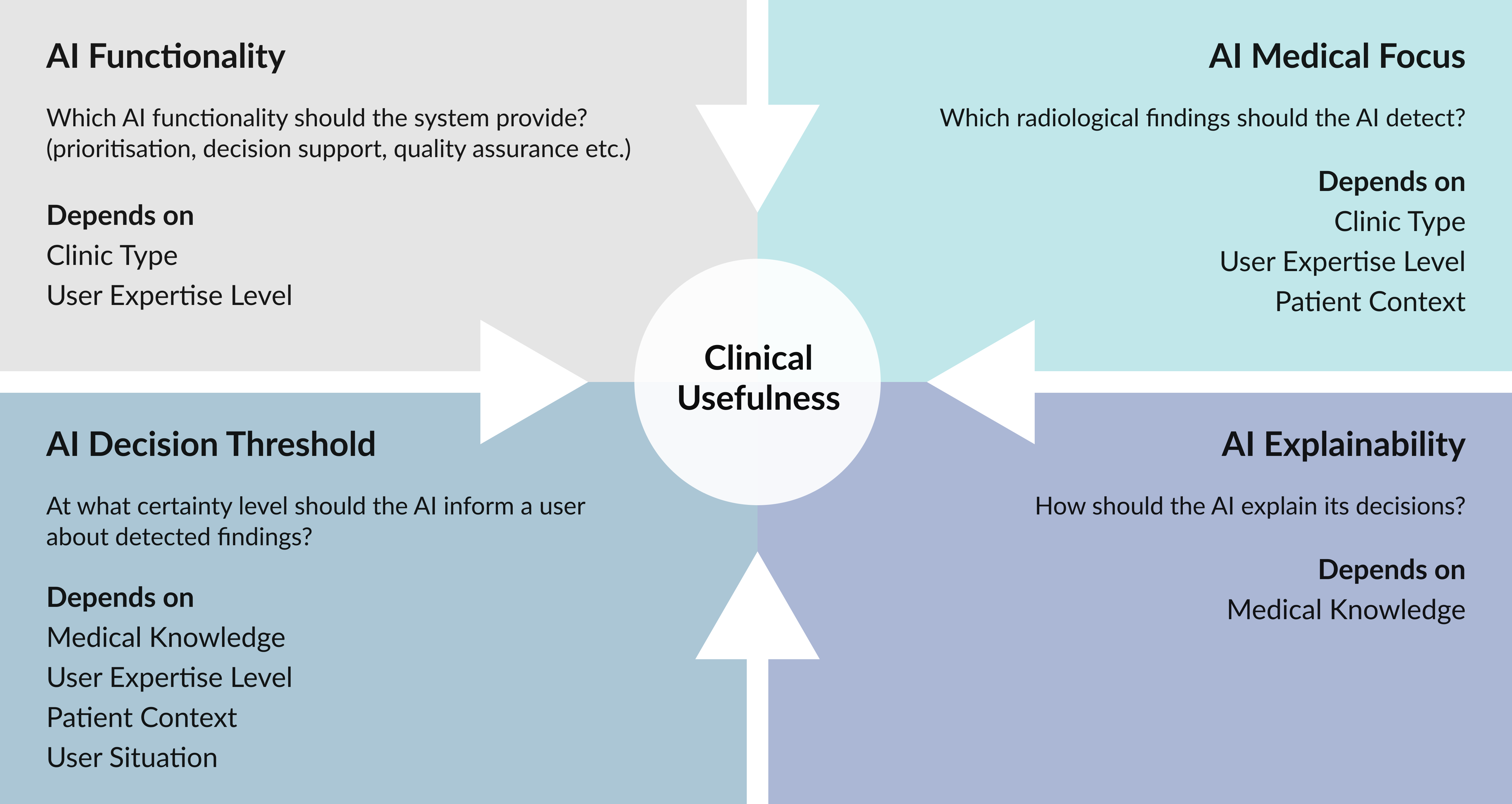}
    \caption{Configuration matrix for achieving clinical usefulness of AI. Showing how the technical AI dimensions need to be configured against the social dimensions, which they depend on.}
    \Description{A diagram showing how different aspects of artificial intelligence (AI) can be configured to achieve clinical usefulness in radiology. The diagram has four categories: AI Functionality, AI Medical Focus, AI Decision Threshold, and AI Explainability. Each category has a question and a list of factors that depend on it. The diagram also has a circle in the centre labelled Clinical Usefulness, indicating the goal of the configuration.}
    \label{fig:dependencies}
\end{figure*}

We identified ten dependencies that emerge from the social dimensions of clinical AI and condition the configuration of four technical dimensions of clinical AI for radiology (see Fig. \ref{fig:dependencies}). Each of the technical dimensions needs to be configured in relation to the local clinical context to achieve clinical usefulness. In this section, we will briefly explain the social dimensions of clinical AI to then explore in-depth the conceptualised dependencies.

\subsection{Social Dimensions of Clinical AI}

\paragraph{Medical knowledge.} This dimension includes concepts and definitions relevant to the medical domain addressed by the innovated AI-based system, for example, the meaning of radiological findings detected by our AI-based system. Familiarity with them supports meaningful collaboration between designers, developers, and medical professionals and reduces the risk of incorrect assumptions throughout the innovation process.

\paragraph{Clinic type.} This social dimension addresses types of clinical sites. Imaging clinics, general hospitals, and specialised hospitals provide unique healthcare services and, thus, cater to the needs of patients with different conditions. Moreover, the type of clinical site determines the available resources, the speciality of medical professionals working there, their workflows, and their goals.

\paragraph{User expertise level.} All medical professionals have different domain expertise. This is evident when comparing junior to senior medical professionals. However, it was also observed between board-certified radiologists. The level of expertise also determines the workload and clinical responsibilities.

\paragraph{Patient context.} This context encompasses the current location of a patient (in or out of a hospital) and their medical history. Patients are the centre of medical work. Their health and well-being are the priority. Thus, by extension, any system supporting healthcare professionals should support patients and depend on their context. 

\paragraph{User situation.} This dimension pertains to the workload, available time, and resources of medical professionals. While the other four dependencies describe relatively stable medical practice, situational context introduces a temporal factor to the work done and may affect the priorities of medical professionals.

\subsection{AI Functionality} \label{sec:functionality}
\begin{figure}
    \centering
    \includegraphics[width=\linewidth]{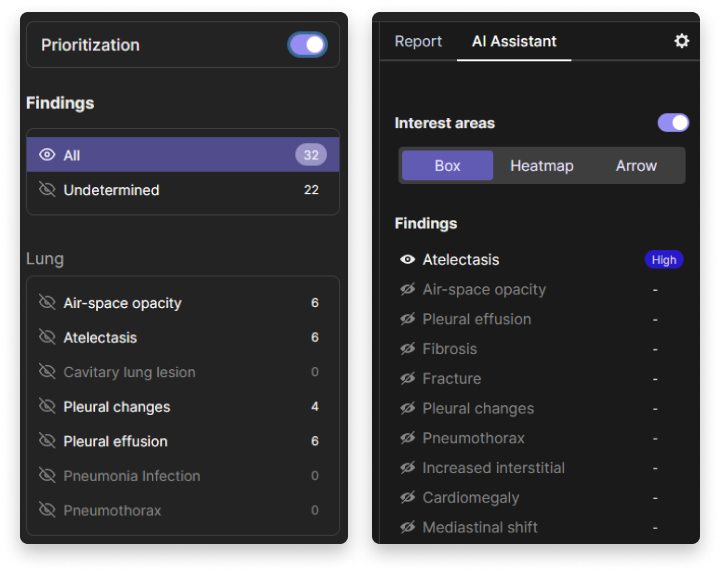}
    \caption{Two UI elements allowing the user to toggle prioritisation and decision support functionalities and to select findings on the X-ray work list and X-ray viewer.}
    \Description{A screenshot of a user interface for a medical AI assistant, displaying prioritised findings related to lung conditions and interest areas in a report format. The left panel lists various lung conditions with associated numbers, and the right panel shows more details and an indicator of high severity for one of the conditions. The user can toggle between different interest areas such as box, heat map, and arrow.}
    \label{fig:functionality}
\end{figure}

Which AI functionality should the system provide? Answering this question defines this technical dimension. The functionalities explored during design interventions (prioritisation and decision support, see, Fig. \ref{fig:functionality}) were linked to the AI model developed for the project this study was a part of. We explored the conditions for these functionalities to provide clinical value and propose a third functionality: quality assurance, which originated during the design interventions. 

\paragraph{Dependency 1: AI functionality depends on clinic type.} Each clinical site has different (1) positions within the healthcare system, (2) amounts of resources available, and (3) workloads related to the size of a clinic. This is why it is important to ensure that AI functionality is implemented in a way that makes sense for the clinical site in which it will be deployed.
 
First, while every radiologist puts the well-being of their patients first, the healthcare systems that they are a part of operate under different incentives. Public and private clinics face different challenges and may require adjusted AI functionalities, for example, \textit{The number of cases in court, medical and legal cases, is way more than what you would get in the public sector. So from the medical director's office [point of view], they would want .... any small thing to be flagged so that we don't get into problems later...  it would be different in K5, compared to the public sector, where even if you missed this, people are rarely taken to court but in a private setting... if they [K5 administration] were to purchase this software, they would insist that it's set... to catch it all but you know of course this would irritate some radiologists}  [S18, Senior, Big general hospital, Kenya]. The difference in the prevalence of legal litigation against medical professionals in public and private healthcare centres highlighted by P05 may run along different axes in other countries. However, it is imperative for the creators of AI-based support systems to envision alternative motivations for the use of their systems and allow appropriate configuration. 

Second, different clinical sites have different financial resources available. This factor, which has rarely been considered during the design of clinical AI-based systems, is a very real limitation for which functionalities will be considered worth the investment. \textit{What is the harm in having a second opinion for each and every case? ...  What is the cost? Is it a cost implication that we have to choose which images to prioritise or what?} [S18, Senior, Big general hospital, Kenya]. The different business models implemented may be detrimental to the usefulness of a system in practice. Providing decision support on all the examinations and detecting all the radiological findings may be too costly for clinics that could use such support the most, e.g., rural hospitals suffering from the lack of qualified radiologists. 

Third, the clinical usefulness of AI functionalities may vary depending on the size of a clinical site, as recounted by a senior radiologist from a busy specialised hospital, \textit{For me, the most relevant aspect of it is triage [prioritisation], but if I have five X-rays to report, then I'm not too worried because I'll get to the 5$^{th}$ X-ray in 20 minutes. But if I have 100 X-rays to go through, I don't want to get to the 100$^{th}$ X-ray and see that it was the one with critical findings. So in a setup where you're not very busy, I don't think it would be very useful} [S19, Senior, Specialised hospital, Kenya]. Conversely, in smaller clinics that serve mostly outpatients, implementing AI that provides quality assurance functionality would provide more value to both the radiologists and the patients. For example, \textit{If I look at it [an examination] and [my colleague] looks at it, no one looks at it until the patient comes back four weeks later, two years later... and then "Oh, look! That's the damn thing." [e.g., a missed tumour] It could be very nice to have this second opinion} [I15, Senior, Imaging clinic, Denmark]. In this imaging clinic, radiologists, rather than being afraid of not reaching a critical patient in time, are worried about missing a critical but subtle finding, e.g., a small nodule, which may signify cancer. This means that the same AI functionality may provide useful support depending on the size of a clinic.

\paragraph{Dependency 2: AI functionality depends on user expertise level.} The value of support in detecting findings on a medical examination decreases with increasing experience. Instead, the assigned workload increases with seniority. Thus, prioritisation and quality assurance functionalities gain importance.

Radiological AI-based decision support typically presents a list of findings detected on an examination accompanied by an XAI visualisation, as also explored in our prototypes. While this mode of support seems straightforward, it misses the reality of clinical practice. Senior radiologists spend a very short time interpreting chest X-rays. To ask them to revisit every examination to discern the validity of AI predictions is wishful. However, when discussing the potential value of AI-based decision support, they focused on quality assurance. Thus, AI should be treated not as an all-knowing peer who is going to point out every finding on an examination but as a safety net that activates only in time of need. For example, \textit{It could read the text we write and say: "Oh, you missed that." That could be good} [I11, Senior, Imaging clinic, Denmark]. This way, the envisioned system would not require the mental effort and time to discern AI output but would inform a radiologist about potentially missed findings based on the report they were writing.

On the other hand, junior radiologists in clinical settings usually take significantly more time to report every examination. Moreover, all of their reports have to be confirmed by a senior colleague. For them, reporting serves as a primary learning exercise. In this context, they envisioned using AI support not as a quality assurance but as a new source of information used to draw their own conclusions. \textit{I would take a look at a chest X-ray, formulate my opinion, and then see what the AI says... If it agrees... good, if it disagrees or finds something that I hadn't, I'll examine it critically... I like getting almost overwhelmed by data, and I sort it out afterwards...} [S14, Junior, Specialised hospital, Denmark]. These two perspectives highlight how workflow, workload, and the act of detecting findings on a medical examination changes with expertise. The educational value created for junior radiologists by verbose explanations of AI's predictions may become a burden for senior radiologists who expect minimal disruption to their existing workflows.\\

\noindent
\fbox{%
    \parbox{0.9\linewidth}{%
        \#1 Recommendation:\\\textbf{\recFunctionality{.}}
    }%
}

\subsection{AI Medical Focus} \label{sec:findings}
\begin{figure}
    \centering
    \includegraphics[width=\linewidth]{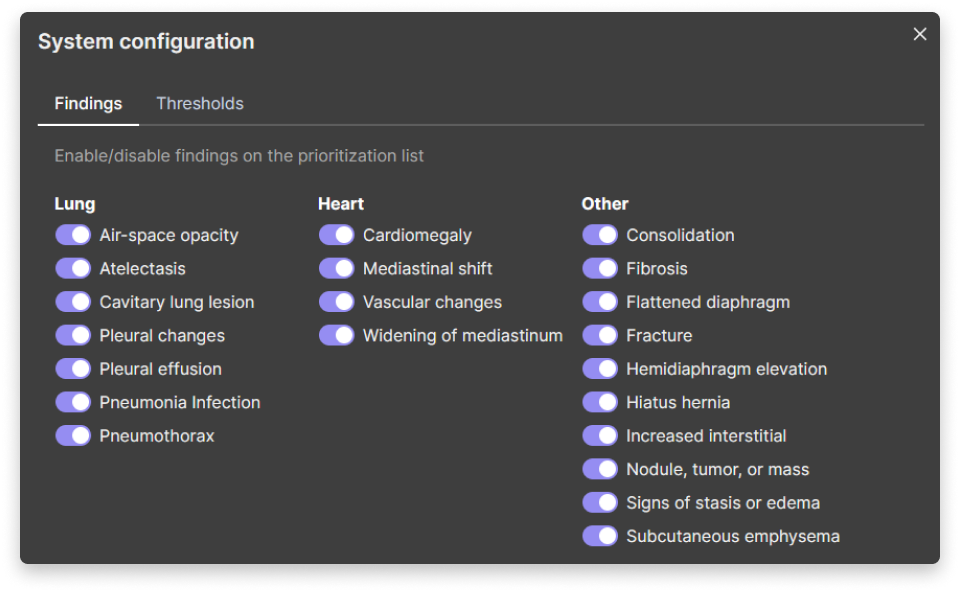}
    \caption{A configuration panel allowing users to select radiological findings detected by the AI model.}
    \Description{A screenshot of a system configuration interface for medical software that allows the user to enable or disable various findings and thresholds related to the lung, heart, and other health conditions. The interface has three columns labelled “Lung”, “Heart”, and “Other”, each with a list of options and toggle switches. Some of the options are alveolar opacity, cardiomegaly, consolidation, and fracture. The interface has a dark theme with white text.}
    \label{fig:focus}
\end{figure}

Which radiological findings should the AI detect? This is where our participants, for the first time, responded, starting with "It depends..." (see Fig. \ref{fig:focus}). Let's explore how to ensure the detected findings are clinically useful in the real world.

\paragraph{Dependency 3: AI medical focus depends on clinic type.} Different clinics take care of different types of patients suffering from different conditions. Types of patients seen in different clinical settings result in a local prevalence of observed radiological findings. As a result, a single fit-them-all system that detects an arbitrarily selected set of findings is not going to provide a similar quality of support across the different clinical contexts.

Imaging clinics and general hospitals usually examine patients referred by general practitioners. Of such patients the majority of the examinations are deemed "normal" or with findings related to infections. Hospitals with emergency departments may observe an increased prevalence of trauma-related findings, whereas specialised hospitals of post-operative, oncological, and chronic nature, as exemplified in these quotes: \textit{That depends on the setting. If you're in a private clinic, most of the X-rays are normal... [I11] If there's something wrong, that could be pneumonia or a tumour, but usually, it’s pneumonia when you go out to a private clinic} [I15, Senior, Specialised Hospital / Imaging Clinic, Denmark]. However, detecting pneumonia would not bring value in other types of settings as highlighted by P05: \textit{If you're working in a trauma centre, the number of critical findings [e.g. pneumothorax or hemothorax] would definitely be more than in [K5], where most of the time it's just coughs and fever} [S18, Senior, Big General Hospital, Kenya]. We argue that to deliver a clinically useful AI-based system for radiologists, it is imperative to understand the local population served by the clinical site where the system is implemented. Otherwise, the developers may risk deploying a system detecting findings that may be objectively relevant to patient management yet not prevalent at the deployment site.

\paragraph{Dependency 4: AI medical focus depends on user expertise level.} Junior radiologists may interpret a single X-ray for up to tens of minutes. Whereas, according to our senior participants, interpreting a chest X-ray takes around 1 to 2 minutes. This means that with experience, many findings become "obvious" and are no feat to detect. When discussing the decision support functionality of our prototypes and previous systems that our participants had piloted, the common complaint related to the detection of "obvious" radiological findings, which took additional time to discern.

\textit{If it's an obvious finding, we'll see that one quickly, and we all agree on it. The problem comes when it's something more subtle} [I06, Senior, Big General Hospital, Kenya]. Detecting the difficult or "subtle" radiological findings is where the value lies for senior radiologists. However, the less experienced, the more support a radiologist may accept. This was captured by P01: \textit{Maybe it'll help the resident radiologist in the first or second year, but I don't think it will help a specialised radiologist with experience because once we can have a look, we can't miss something like this} [I01, Senior, Small General Hospital, Kenya]. This means that in order to support different radiologists in practice, AI-based systems may need to allow users to select findings to receive support with. Without such configuration, discerning AI predictions regarding "obvious" findings, even when true, would result in more time spent and annoyance. 

\paragraph{Dependency 5: AI medical focus depends on patient context.} Radiologists are not interpreting medical imaging to find every possible finding. Rather, they are interpreting them to help the ordering clinicians take action in patient management. Such actions usually occur when a new condition is being diagnosed, or a patient's health may be at risk. However, the clinical meaning of certain radiological findings depends on the location of a patient. This means that a finding may be expected when observed in an examination of a patient who is admitted to a hospital. Whereas the same finding observed in an examination of a patient who is not admitted to a hospital may warrant immediate action.

Our participants stressed that useful prioritisation should consider patients' medical history to filter out already-known findings, which our prototype could not do. \textit{ But how urgent is it? We know that pneumothorax has decreased. It's a big heart, but it's much smaller than it was a week ago. It has [pleural] effusion, but much, much less than it was a week ago. That's the thing we miss with this} [I11, Senior, Specialised Hospital / Imaging Clinic, Denmark]. In this quote, P11 explains that the examination they looked at may not be urgent at all despite the fact that the AI correctly detected three findings, one of them (pneumothorax) being life-threatening. These findings would not be urgent if they were already known to the ordering clinician. In such a case, the patient would have already been undergoing treatment, and this examination's sole purpose was to control its progress. In specialised hospitals and bigger general hospitals, patients often have taken several X-rays to monitor the progress of treatment. This means that the same findings, but of different severity, will be visible on their examinations. The ability to assess the detected findings in the light of patient history is crucial to correctly prioritise findings that warrant clinical action.

When looking at radiologists' work from the perspective of contributing to the broader clinical work, it is counterproductive to prioritise findings that clinicians taking care of a patient are already aware of. In other words, a radiological finding may be relevant to detect on examinations from patients who are not admitted to a hospital, but not so much for patients currently admitted. A senior radiologist explained, \textit{It depends on the findings, and it depends on the patient... some findings in the out-patients would be more important to be prioritised than if they're in-house. Because if they're in-house, then I would suspect that someone not from the radiology department would have looked at them. If it's out-patient, then nobody has looked at them...} [I10, Senior, Specialised Hospital, Denmark]. Whereas, as explained by P10, patients referred from outside of a hospital are more likely to have conditions that their doctors are unaware of. Thus, the location of the patient is crucial to selecting which findings are relevant to receiving support from an AI-based system. \\

\noindent
\fbox{%
    \parbox{0.9\linewidth}{%
        \#2 Recommendation:\\\textbf{\recFocus{.}} 
    }%
}

\subsection{AI Decision Threshold} 
\begin{figure}
    \centering
    \includegraphics[width=\linewidth]{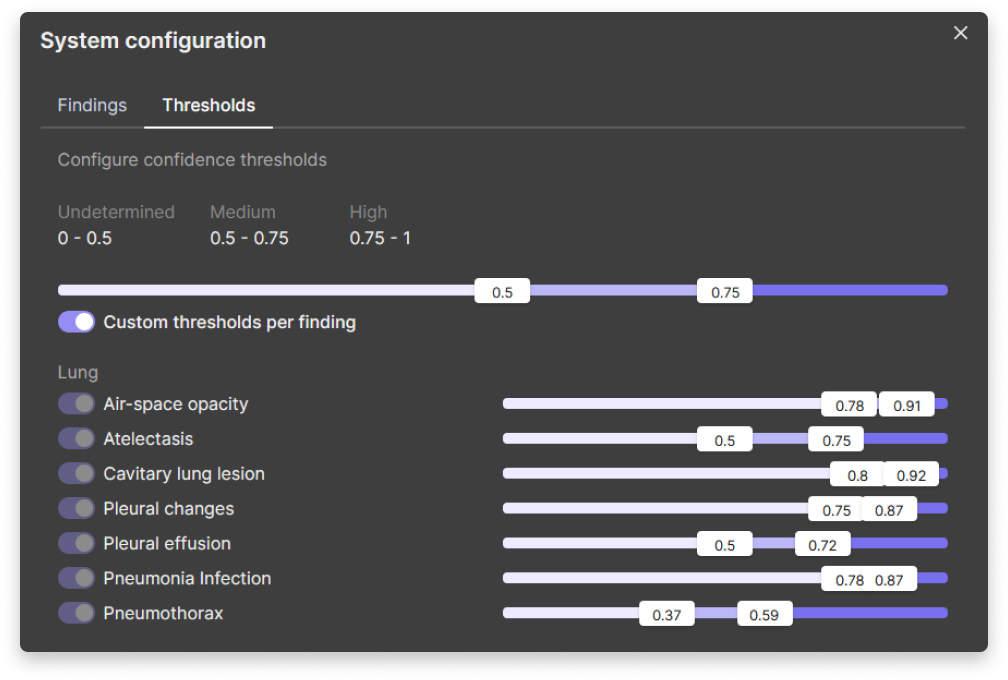}
    \caption{A configuration panel allowing users to select AI decision threshold globally and per finding. In this version, we introduced two levels for quick access depending on the user situation: high confidence and medium confidence.}
    \Description{A screenshot of a system configuration window for setting AI confidence thresholds for various medical findings related to lung conditions. The window has two tabs, “Findings” and “Thresholds”, and allows the user to adjust the threshold values for undetermined, medium, and high confidence levels. The window also has sliders for six different types of lung findings: Lung, Airspace opacity, Cavitary lung lesion, Pleural changes, Pleural effusion, Pneumonia infection, and Pneumothorax. Each finding has three coloured bars representing the confidence levels with numerical values.}
    \label{fig:threshold}
\end{figure}

At what certainty level should the AI inform a user about detected findings? Specifying when a radiological AI-based system should inform a user about a finding is usually done by specifying a decision threshold (see Fig. \ref{fig:threshold}). Selecting a specific threshold value determines the measured performance of an AI model captured by evaluation metrics like specificity, sensitivity, or positive and negative predictive values. Arguably, in practice, a decontextualised performance value is less important than the practical consequences of selecting a specific threshold level. Every time an AI model detects a finding (based on a selected threshold), a radiologist may have to take action to assess it. The balance between clinical value and additional burden is thus closely tied to how well the threshold is configured to match the local clinical context. We conceptualised four dependencies that influence the configuration of the AI decision threshold.

\paragraph{Dependency 6: AI decision threshold depends on medical knowledge.} While some of the radiological findings are well understood across the contexts, some definitions are more subjective and their meanings change across countries. Infiltration or consolidation are two examples of radiological findings which have been found to be used differently in clinical practice in Denmark and Kenya. Moreover, some of the findings were too vague for the radiologists to decide how to assess them, for example, \textit{I think vascular changes can mean one of two things if it's the big vessels - I think it's important to have it, e.g., if the computer can say the aorta is big... It could also be about... the small vessels, and then it's more like stasis. Then it's quite different} [I17, Junior, Specialised hospital, Denmark]. The underlying definition of a finding, in this described case, affected how P13 understood the condition and what threshold level they deemed appropriate. Within radiology, chest X-rays are a particularly subjective modality. Due to their visual complexity, radiologists rely on their expertise to interpret the observed findings. Precise definitions used to label data the AI model was trained on are crucial to assess the predictions.

\paragraph{Dependency 7: AI decision threshold depends on user expertise level.} Often, junior and senior radiologists are juxtaposed as two groups of AI support end-users with different needs. This is also visible in the strategy for selecting thresholds. 

When used by junior radiologists, both junior and senior radiologists (who supervised them) leaned towards accepting AI predictions only with a high degree of certainty. As explained before, the interpretation of chest X-rays is uniquely subjective. It takes experience to report them with a high degree of certainty. In this context, an uncertain AI prediction would jeopardise the learning process and introduce more confusion, resulting in more work for the junior students and their supervisors. 

On the other hand, allowing senior radiologists to set the threshold for different findings according to their personal preferences could entice them to utilise the system in their own way. P05, who also had senior administrative experience, explained that senior radiologists do not always have the same level of expertise and may need different levels of support. \textit{This would be amazing. I wouldn't want to do it [adjust threshold] at the administrative level because ... not all the radiologists in the department have the same capabilities. So I'd rather let people set it for themselves} [S18, Senior, Big general hospital, Kenya]. By enabling users to select the AI decision threshold on their own, they could build trust by incrementally including AI in their own practice.

\paragraph{Dependency 8: AI decision threshold depends on patient context.} Diverging from a fixed threshold level defined at a system level towards finding-level threshold specification may boost the clinical usefulness of AI-based systems for radiology. A finding-level threshold specification would allow radiologists to stratify which findings in a given context are more relevant. 

They could do it by lowering the threshold. A lower threshold would be associated with a higher rate of false positive prediction for that particular radiological finding. Thus, more work for radiologists. However, for a subset of findings, our participants were willing to accept more false positive detections if it would benefit their patients. \textit{For pneumothorax, I would probably lower the threshold because you would want to find every pneumothorax there is. But for some other stuff, like fibrosis, I would probably have a higher threshold because that's not critical} [S13, Senior, Specialised hospital, Denmark]. Based on design interventions with the third prototype, our participants saw a utility in such fine-grained configuration. \textit{I think the relevance of certainty [threshold] is the clinical implication of the diagnosis. So, something like a pneumothorax needs some form of intervention... whereas on a suspected infection, a clinician may go ahead and treat it even if the X-ray is normal. So that's why it may not be such a big deal whether I call a pneumonia or not. Whereas a pneumothorax might need a chest tube insertion. It's a do-or-die call} [S19, Senior, Specialised hospital, Kenya]. As shown in this quote, the clinical implications for a patient made radiologists more accepting of false positives. Meanwhile, findings that were less severe or that could be discerned using other indicators, e.g., clinical indicators (cough, fever) to decide on pneumonia diagnosis, were less preferred to lower the threshold. This suggests that configuring AI decision threshold on a finding level could reduce the workload associated with false positive predictions and help focus AI support.

\paragraph{Dependency 9: AI decision threshold depends on user situation.} Radiologists' approach to AI support changes with time. In this paper, we uncovered two temporal aspects that affected how radiologists thought of configuring AI decision thresholds: the time spent using the system and the rhythm of clinical work.

One of the common comments when discussing the threshold with our participants was about its arbitrary nature. Radiologists wondered what the real-life consequences of changing the threshold would be. Based on these concerns, our final prototype included an estimate of false positive predictions. These values, while more relatable, were still considered difficult to imagine in real practice both for senior and junior radiologists. \textit{I mean, it's a bit arbitrary at this moment because you don't have any idea what the effect is} [I17, Junior, Specialised hospital, Denmark]. \textit{It would be nice to be able to adjust this... try all this out and see in real life how many cases it's missing or over-calling} [S19, Senior, Specialised hospital, Kenya]. These quotes highlight that such essential development tasks as selecting a threshold have little to no basis in clinical practice. They uncover a need for a better translation between the domains of AI and Health to support meaningful configuration. Currently, this translation has to be conducted through real-world experimentation in the final context of use. This way, medical professionals may gain a practical understanding of what the changes to the threshold mean and further purposefully and consciously adjust it to fit their work.

The second temporal aspect of selecting an appropriate threshold relates to the routine of end users. Radiologists saw an advantage in adjusting the threshold depending on their workload. For example, a specialised radiologist from a busy specialised hospital mentioned, on \textit{ Fridays, we tend to be more active because if you leave a long list on Friday, the turnaround time will be way longer - there is very low coverage over the weekend [few on-call doctors]... and then Monday tends to be very busy} [I04, Senior, Specialised hospital, Kenya]. During this conversation, the radiologist concluded that lowering the threshold could help them ensure that no examinations with critical findings were left to be reported after the weekend. These two aspects highlight that what radiologists consider a useful level of detection (including false positive predictions) may vary throughout the use.\\

\noindent
\fbox{%
    \parbox{0.9\linewidth}{%
        \#3 Recommendation:\\\textbf{\recThreshold{.}}
    }%
}

\begin{figure*}[h]
    \centering
    \includegraphics[width=\linewidth]{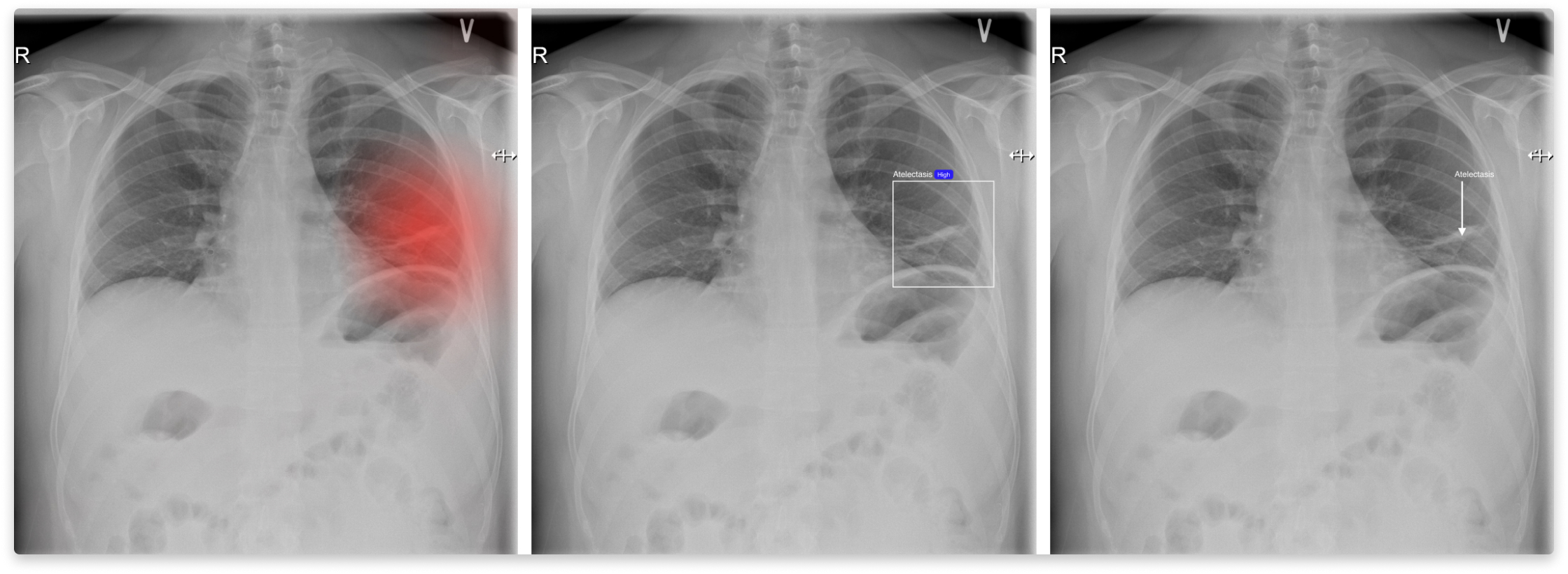}
    \caption{Different XAI methods available in the prototype. From left to right: gradient overlay, bounding box, arrow.}
    \Description{A series of three X-ray images showing the human chest, with annotations highlighting specific areas of interest in the lungs. The first image has a red highlighted area on one of the lungs, the second image has a blue rectangular box around the same area, and the third image has an arrow pointing to that area.}
    \label{fig:xai-methods}
\end{figure*}

\subsection{AI Explainability} \label{sec:xai}
How should the AI explain its decisions? Understanding AI predictions supports building trust towards AI-based systems. In this study, we explored three visual ways of explaining AI predictions: heat maps, bounding boxes, and arrows (see Fig. \ref{fig:xai-methods}). We discovered that no single method can support the explainability of all the radiological findings.

\paragraph{Dependency 10: AI explainability depends on medical knowledge.} The visual appearance of radiological findings dictates the best way to highlight them for radiologists for inspection. Radiologists discern between different radiological findings based on their visual appearance. Their presentation ranges from barely visible nodules to diffused opacities (areas of less transparency) present across both lungs. The breadth of visual impressions suggests the need for flexibility, \textit{I think that both ways of displaying the findings are fine, but for different pathologies. I mean, the heat map makes sense in this case for pneumothorax because it's a very extensive finding. And for the fracture, it makes sense to see it with a box, whereas the heat map doesn't make that much sense. It becomes too blurry...} [S13, Senior, Specialised hospital, Denmark]. Radiologists preferred bounding boxes for more contained findings, whereas the more diffused, the more inclined they were towards the heat map. An important factor when designing XAI for chest X-rays is allowing for inspection of the underlying examination. The main purpose of XAI is to direct radiologists' attention to the detected findings. To assess the validity of a prediction, radiologists have to inspect the examination itself without additional overlays.\\

\noindent
\fbox{%
    \parbox{0.9\linewidth}{%
        \#4 Recommendation:\\\textbf{\recXAI{.}}
    }%
}

\section{Discussion}
\begin{figure*}[h]
    \centering
    \includegraphics[width=0.85\textwidth]{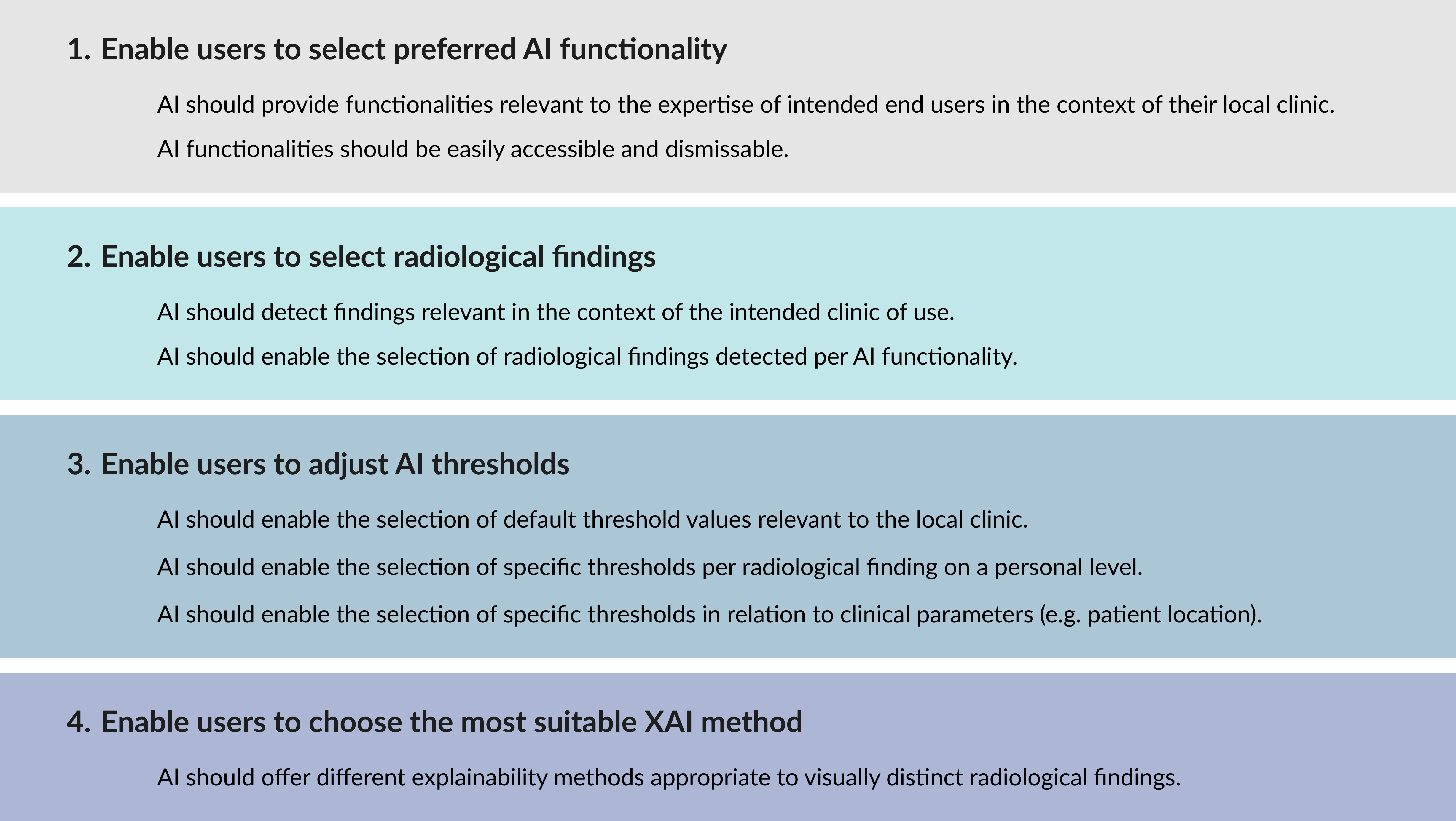}
    \caption{Four design recommendations on how to achieve clinically useful AI-based systems. Accompanied by more in-depth considerations.}
    \label{fig:recommendations}
    \Description{An infographic showing how users can customise AI for clinical usefulness, with four categories of options: AI functionality, AI medical focus, AI decision threshold, and AI explainability. Each category has a coloured bar with white or black text explaining the choices. The infographic has a light background and a title that says: Towards clinical usefulness: Configurable AI.}
\end{figure*}

In this paper, we investigated how to design AI for clinical usefulness in different clinical contexts of radiology practice. Based on extended research through design study, we provided four practical recommendations on addressing ten dependencies emerging from the social dimensions of clinical practice (Fig. \ref{fig:recommendations}).  By engaging radiologists in two different countries from the Global North and Global South, respectively, we found that the radiologists’ practices in Kenya and Denmark were rather similar, possibly due to resemblances in medical education, scientific models, and ethics. The social dimensions derived from this study therefore orient towards similarities that cut across country specifics. However, the types of healthcare systems and present IT infrastructures differed quite substantially and need to be accounted for during AI innovation, which goes beyond this study. In this section, we will discuss how these recommendations may be enacted during the innovation process of clinical AI for different clinical contexts. 

\subsection{\recFunctionality{}}
Configuring clinical AI-based support systems to suit local environments is essential, as one-size-fits-all approaches often fail to address their unique needs \cite{Gu2021LessonsPathologists, Jacobs2021DesigningLens, Yang2019, Pine2020RightPractice}. In this study, we discovered how social dimensions of clinical practice condition what kind of AI functionality is considered useful. We argue that for AI to match local requirements, it needs to be configured throughout the innovation process with the intended context of use in mind, i.e., the expertise of the end-users and the work performed in their clinical site.  This is especially relevant, as clinical AI is often afflicted by the problem of late realisation \cite{Yang2018InvestigatingLearning, Girardin2017WhenScientists}.

When addressing expertise-related needs for support, previous research in radiology showed that AI-based systems have different effects on junior and senior radiologists \cite{Zhou2024InterpretableImages}. Even more, Tong et al. \cite{Tong2023IntegrationManagement} investigated two strategies, what they called "optimised" and "all-AI", for AI support of junior and senior radiologists in thyroid nodule management. They reported that the best results were obtained when the type of support was configured to the expertise of the radiologist. However, our study showed that personal preferences play a deciding factor only if the AI functionality is appropriate in the context of the local clinic. Selecting the best way to prioritise findings will not make sense if there are only a few examinations to prioritise to begin with. AI-based systems should be designed to respond to fit the utility gap in a clinic and then be configured to the varying needs and preferences of different end-users, depending on their level of experience, knowledge, and confidence. 

The personal configuration of functionality also captures the integration - a famously difficult task when innovating clinical AI \cite{Yang2016InvestigatingHelp, Pine2020RightPractice, Thieme2023DesigningTreatment}. Many AI-based systems fail in practice due to providing support at the wrong time \cite{Benda2020, Hollander2004EffectsPain, Ginestra2019ClinicianShock, Cai2019Human-CenteredDecision-Making}. Some of the AI integrations introduce a new step in the practice. A step that sometimes cannot be skipped \cite{Benda2020}. This study shows, seconding previous research, that the integration of AI into work practices has to be flexible \cite{Bossen2023BatmanSpecialists}. Clinical work is always changing, and so are the needs for AI support. Thus, we recommend that clinical end-users should be in control of which AI functionalities are a part of their current routine.

\subsection{\recFocus{}}
The innovation of AI-based systems is often initiated and defined by technical opportunities, e.g., access to medical data \cite{Shneiderman2020Human-CenteredTrustworthy, Shneiderman2020Human-CenteredIdeas, Xu2019TowardInteraction, Zheng2017Hybrid-augmentedCognition}. As such, the medical and social aspects of the systems are sometimes addressed only after the technology has gone through several rounds of development \cite{Amershi2019}. This inadvertently means that certain assumptions about the medical focus are made \cite{Zajac2023GroundAI}. Present radiological AI models tend to detect findings relevant to the local radiologists involved in the data creation process \cite{Nguyen2020, Irvin2019, Wang2019}. However, we showed that the prevalence and clinical meaning of radiological findings varies based on the clinic type and patient context. This affects the usefulness of clinical systems in different settings and their transferability \cite{Nolan1991FactorsSystems, Zajac2023GroundAI}. Thus, it is critical to investigate the intended clinical context of use prior to deciding on the medical focus of the AI-based system and to allow medical professionals to set the scope of support relevant to them and their practice.

Moreover, the clinical meaning of radiological findings is tied to the patient context and not only the type of medical condition, i.e., a radiological finding expected in an in-patient examination can be life-threatening when found in an out-patient one. This discovery deepens our understanding of how medical professionals make decisions and in what situations they may need AI support. This finding contrast with systems where certain radiological findings are consistently considered urgent in patient care \cite{Xie2020CheXplain:Analysis}. We suggest that linking clinical information about a patient with detected findings may better reflect radiologists' actual decision-making practices and result in improved usefulness of the AI-based system. This is why we recommend including clinical information in conjunction with AI predictions to better respond to the real-world needs of medical professionals.

\subsection{\recThreshold{}}
Selecting AI decision threshold has significant ethical \cite{Birch2022ClinicalValues}, performance \cite{Sahoo2021ReliableCalibration}, and clinical \cite{Vistisen2022ArtificialSetting} consequences for AI-based systems, and it has been a notable research topic in the AI and Health communities. Recently, it gained footing in the HCI design community. Kocielnik et al. \cite{Kocielnik2019WillAI} explored how the decision threshold affects the number of false positive and false negative predictions, significantly altering users system perception. While from the technical point of view, the accuracy may be the same, the distribution of false positives and false negatives may have severe clinical consequences. Our participants warned that false positive predictions require additional time and resources to discern and that the potential benefits of AI often do not justify this additional cost, resulting in the failure of the AI-based systems in clinical practice \cite{Thieme2020MachineHealth, Baxter2020BarriersReadmissions, Matthiesen2021ClinicianStudy, Bach2023IfSupport}.

However, until AI reaches 100\% accuracy, false positive predictions are the reality of AI-based systems. Improving performance is only one way of addressing them. In this paper, we offer another outlook, namely, addressing the cost-benefit ratio of AI predictions. This ratio is not static. Just like clinical practice, it fluctuates and depends on time, workload, known critical cases, and available resources. In certain situations, medical professionals may accept more false positive predictions, e.g. when making sure that there are no critical findings in a queue of examinations that will not be looked at over the weekend. This means that regardless of how well an AI decision threshold is preset, AI will not provide the same value throughout its use in clinical practice. Supporting end-users in configuring the AI decision threshold depending on their local needs can improve the clinical usefulness of AI-based systems. Thus, designers and developers should enable end-user configuration of decision thresholds in clinician-facing AI systems.

\subsection{\recXAI{}}
It has been long established that explainable AI fosters trust and increases the usefulness of the predictions \cite{Dhanorkar2021WhoLifecycle, Jacobs2021DesigningLens}. Especially in the healthcare domain, the reasoning and explanations are sometimes more valuable to end users than the predictions themselves \cite{Matthiesen2021ClinicianStudy, Cai2019Human-CenteredDecision-Making} or can lead to envisioning new ways of using an AI-based system altogether \cite{Jin2020}. However, simply revealing the decision-making process of machines to humans is not enough to provide useful explanations \cite{Morrison2018}. Instead, our study suggests that for XAI methods to be effective in explaining medical conditions, they must be configured to how medical professionals assess those conditions. This means that even proven methods used in medical imaging, like heat maps or bonding boxes, when used to highlight incompatible conditions, may cause confusion and require additional work to discern. To this end, we recommend that to ensure the clinical usefulness of XAI methods, they should be configurable in accordance with medical knowledge.

\section{Limitations and Future Work}
This work is not without its limitations. As explored in this paper, when interacting with the prototype, radiologists envisioned support functionalities like quality assurance through the assessment of written reports against AI's interpretation of findings on a chest X-ray. This functionality was outside of the prototyped prioritisation and decision support. This choice was dictated by the capabilities of the underlying AI model and the innovation direction of the greater project this study was a part of. We believe that this mismatch perfectly exemplifies the difficulty of innovating clinically useful AI-based systems and motivates further research into a meaningful configuration of AI-based systems, especially at the defining early stages of work. In addition, the study did not take into consideration the differences in clinicians' attitudes and expectations towards AI as well as the collaborative aspects of reporting, which may be worth considering in studies that follow up on designing for clinical usefulness. 

We also acknowledge the limited variability of clinical sites in Denmark compared to the visited sites in Kenya due to difficulties gaining access. While it is a strength of the study that medical professionals from very different countries participated, future studies need to further explore how geographic and cultural differences play out in regard to successfully designing for and transferring AI-based systems to entirely different healthcare systems and IT infrastructures. Moreover, this project commenced before large language models experienced a performance leap. We believe that their ability to parse and produce text may be an opportune, although challenging avenue for AI support to explore \cite{yildirim2024multimodal}.

\section{Conclusions}
Innovating clinical AI-based systems is a challenging task. By investigating design interventions conducted with radiologists across diverse clinical contexts in Denmark and Kenya, we identified four key technical dimensions that require careful configuration: AI functionality, AI medical focus, AI decision threshold, and AI explainability. To support the innovation of clinically useful AI-based systems, we derived four concrete recommendations of what we propose to call "configurable AI" pertaining to the four key technical dimensions. 

Moreover, we explored how dependencies originating from the social dimensions of local clinical practice condition the clinical usefulness of the uncovered technical dimensions. AI functionalities (e.g., prioritisation or decision support) should be configured to provide value in the intended type of clinical site and to match the level of medical expertise of end users. AI medical focus (the detected findings in radiology-focused systems) should be configured in relation to the patient's context, the level of medical expertise of the end-users, and the type of clinical site. The AI decision threshold should be configured according to the medical knowledge (e.g., the clinical meaning of radiological findings), the patient's context, the level of medical expertise of the end users, and the user situation (e.g. time of day). Finally, the explainable AI should be configurable in accordance with medical knowledge to provide maximum value to the end-users. 

Our findings highlight the need for designers and developers to consider these dependencies throughout the innovation process, both before-use and in-use, to ensure that AI-based systems are effectively configured to meet the needs and requirements of their intended clinical contexts. By adhering to these recommendations and considering the dependencies uncovered in our study, designers and developers can contribute to the successful innovation of clinically useful AI-based systems in radiology, ultimately improving patient care and clinical outcomes.

\bibliographystyle{ACM-Reference-Format}

\bibliography{refs_1,refs_2}

\end{document}